\documentclass{article}

\usepackage{arxiv}
\usepackage[utf8]{inputenc} 
\usepackage[T1]{fontenc}    
\usepackage{hyperref}       
\usepackage{url}            
\usepackage{booktabs}       
\usepackage{amsfonts}       
\usepackage{nicefrac}       
\usepackage{microtype}      
\usepackage{lipsum}		    
\usepackage{graphicx}
\usepackage{natbib}
\usepackage{doi}
\usepackage{siunitx}
\usepackage{amsmath,amssymb,amsfonts}
\usepackage{mathtools} 

\usepackage{amssymb}
\usepackage{pifont}
\usepackage{float}
\usepackage{caption}
\usepackage{subcaption}
\usepackage{comment}
\usepackage{mathtools}

\title{Technology-Circuit-Algorithm Tri-Design for Processing-in-Pixel-in-Memory (P$^2$M)}

\author{Md Abdullah-Al Kaiser\thanks{These authors contributed equally to this work.} \\
	University of Southern California\\
	Los Angeles, CA \\
	\texttt{mdabdull@usc.edu} \\
	\And
	Gourav Datta$^*$ \\
	University of Southern California\\
	Los Angeles, CA \\
	\texttt{gdatta@usc.edu} \\
	\And
        Sreetama Sarkar \\
	University of Southern California\\
	Los Angeles, CA  \\
	\texttt{sreetama@usc.edu} \\
	\And
        Souvik Kundu \\
	Intel Labs\\
	San Diego, CA  \\
	\texttt{souvikk.kundu@intel.com} \\
	\And
        Zihan Yin \\
	University of Southern California\\
	Los Angeles, CA  \\
	\texttt{zihanyin@usc.edu} \\
	\And
        Manas Garg \\
	University of Southern California\\
	Los Angeles, CA  \\
	\texttt{manasgar@usc.edu} \\
	\And
	Ajey P. Jacob \\
	Information Sciences Institute\\
	Marina del Rey, CA  \\
	\texttt{ajey@isi.edu} \\
        \And
	Peter A. Beerel \\
	University of Southern California\\
	Los Angeles, CA  \\
	\texttt{pabeerel@usc.edu} \\
 	\And
	Akhilesh R. Jaiswal \\
	University of Southern California\\
	Los Angeles, CA  \\
	\texttt{akhilesh@usc.edu} \\
}

\renewcommand{\shorttitle}{Technology-Circuit-Algorithm Tri-Design}

\begin{document}
\maketitle

\begin{abstract}
The massive amounts of data generated by camera sensors motivate data processing inside pixel arrays, i.e., at the extreme-edge. Several critical developments have fueled recent interest in the processing-in-pixel-in-memory paradigm for a wide range of visual machine intelligence tasks, including (1) advances in 3D integration technology to enable complex processing inside each pixel in a 3D integrated manner while maintaining pixel density, (2) analog processing circuit techniques for massively parallel low-energy in-pixel computations, and (3) algorithmic techniques to mitigate non-idealities associated with analog processing through hardware-aware training schemes. This article presents a comprehensive technology-circuit-algorithm landscape that connects technology capabilities, circuit design strategies, and algorithmic optimizations to power, performance, area, bandwidth reduction, and application-level accuracy metrics. We present our results using a comprehensive co-design framework incorporating hardware and algorithmic optimizations for various complex real-life visual intelligence tasks mapped onto our P$^2$M paradigm.
\end{abstract}



\maketitle

\section{Introduction}

High-resolution and high frame rate cameras generate and transmit a large amount of data to back-end processors/accelerators for today's widespread artificial intelligence (AI)-enabled computer vision (CV) applications \cite{auto_driving, obj_track}. This physical segregation between the sensors and processors creates energy and bandwidth bottlenecks in modern CV pipelines. Researchers have developed approaches to \textit{process and compress} data close to the sensor node instead of performing all-backend processing to reduce the amount of transferable data to mitigate this bottleneck. 

Typically these approaches can be classified into three categories- (1) near-sensor processing, (2) in-sensor processing, and (3) in-pixel processing. In near-sensor processing, a processor/accelerator is placed adjacent to the CMOS image sensor (CIS) chip to improve energy and bandwidth efficiency by reducing the data transfer cost between the sensor chip and the cloud/edge processor \cite{near_sensor_sony, near_in_sensor_survey}. In contrast, the in-sensor approach utilizes an analog or digital signal processor at the periphery of the sensor chip \cite{in_sensor_sleep_spotter, in_sensor_ivs}. Though these systems are more energy and bandwidth efficient than traditional systems, they suffer from data transfer bottlenecks between the sensor and periphery or the back-end processor. On the other hand, the in-pixel processing approach enables early processing by performing computations inside the pixel array, thus minimizing the subsequent data transmission.  

Several in-pixel processing works have been reported in \cite{senputting, scamp_simd, prog_kernel, in_pixel_cifar10}. To achieve high classification accuracy, complex machine learning applications require multi-bit, multi-channel convolution, batch normalization (BN), and Rectified Linear Units (ReLU) operations.  However, approaches in \cite{senputting, scamp_simd} embed binary weights inside the pixel array, \cite{scamp_simd} utilizes area and energy inefficient digital in-pixel hardware, and \cite{senputting, prog_kernel, in_pixel_cifar10} lack multi-channel convolution capabilities. Furthermore, most existing in-pixel approaches focus on simplistic datasets, such as MNIST, \cite{senputting, scamp_simd}, CIFAR-10 \cite{in_pixel_cifar10}, that do not represent realistic computer vision applications. Reference \cite{aps_p2m} proposes a novel multi-bit multi-channel weight-embedded in-pixel processing approach along with BN and ReLU operations and reports 11\si{\times} energy-delay-product (EDP) improvement on visual wake words (VWW) dataset. Moreover, \cite{aps_p2m_detrack} evaluates their in-pixel processing solution on the large-scale BDD100K dataset. However,  weights are implemented utilizing the transistor’s width in \cite{aps_p2m, aps_p2m_detrack}, which is fixed during fabrication. 
Hence, in this work, we extend embedding weights through a hybrid CMOS-RRAM circuit where the weights can be reconfigurable according to different applications.

\begin{figure}[!t]
\centering
\includegraphics[width=0.9\linewidth]{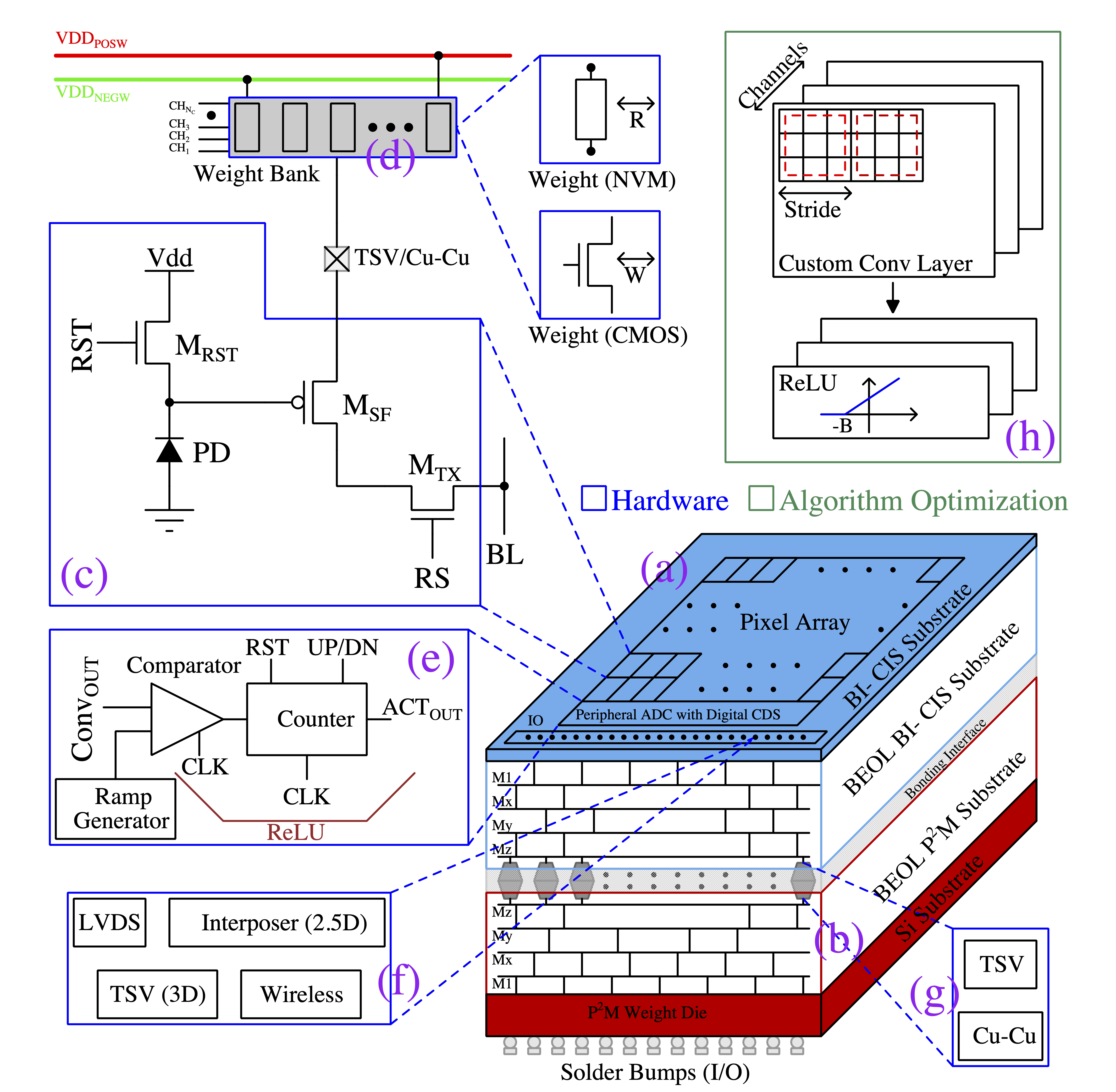}
\caption{Overall P$^2$M-enabled CIS system. (a) Back-side illuminated CMOS image sensor (BI-CIS) die, (b) weight-containing die, (c) pixel circuit, (d) multi-bit multi-channel positive and negative weight banks (mapped into transistor's width (CMOS), or the resistance state (NVM)), (e) SS-ADC performing the ReLU and part of BN operations, (f) IO configurations, (g) different integration technologies for bonding interface, (h) algorithm-hardware co-design framework.}
\label{overall_system}
\end{figure}

There are intricate design trade-offs in the technology (chip integration and IO), circuit, and algorithm layers for the in-pixel processing approaches. Embedding computation, such as matrix-vector multiplication, inside the pixel array can reduce the pixel density. However, utilizing heterogeneous 3D integration technologies, logic or memory substrate can be stacked vertically with CIS. On the other hand, state-of-the-art (SOTA) algorithmic accuracy demands a large number of channels and a small stride in the first few layers that increase the number of weights per pixel and, hence, the area. As a result, it may limit the minimum pixel pitch and, consequently, the resolution of the CIS. Advanced process nodes with lower contacted poly pitch (CPP) and metal pitch (MP) may support the algorithmic requirements; nevertheless, the coarse bond pitch of 3D integration technology can reduce the area benefits of the advanced process nodes. Besides, the in-pixel solution's overall latency and energy consumption depend on the algorithmic parameters (number of channels, kernel, and stride number) and hardware configurations (ADC bit precision, different IO). All the important design and performance parameters are interdependent, and conflicting trade-offs are hard to optimize. This article holistically studies these design and performance trade-offs. 

\begin{figure}[!b]
\centering
\subfloat[]{\includegraphics[width=0.5\linewidth]{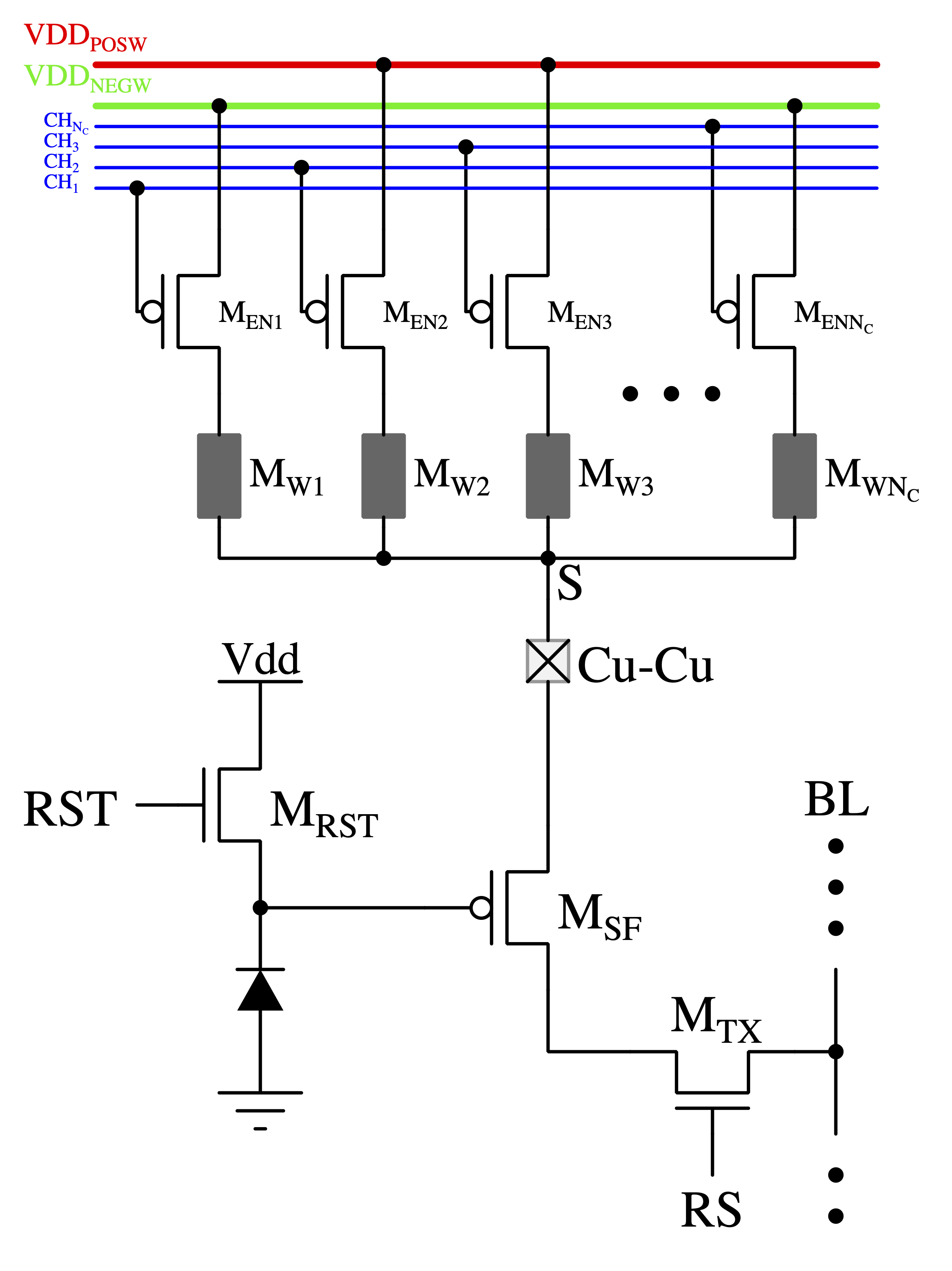}} \\
\subfloat[]{\includegraphics[width=0.5\linewidth]{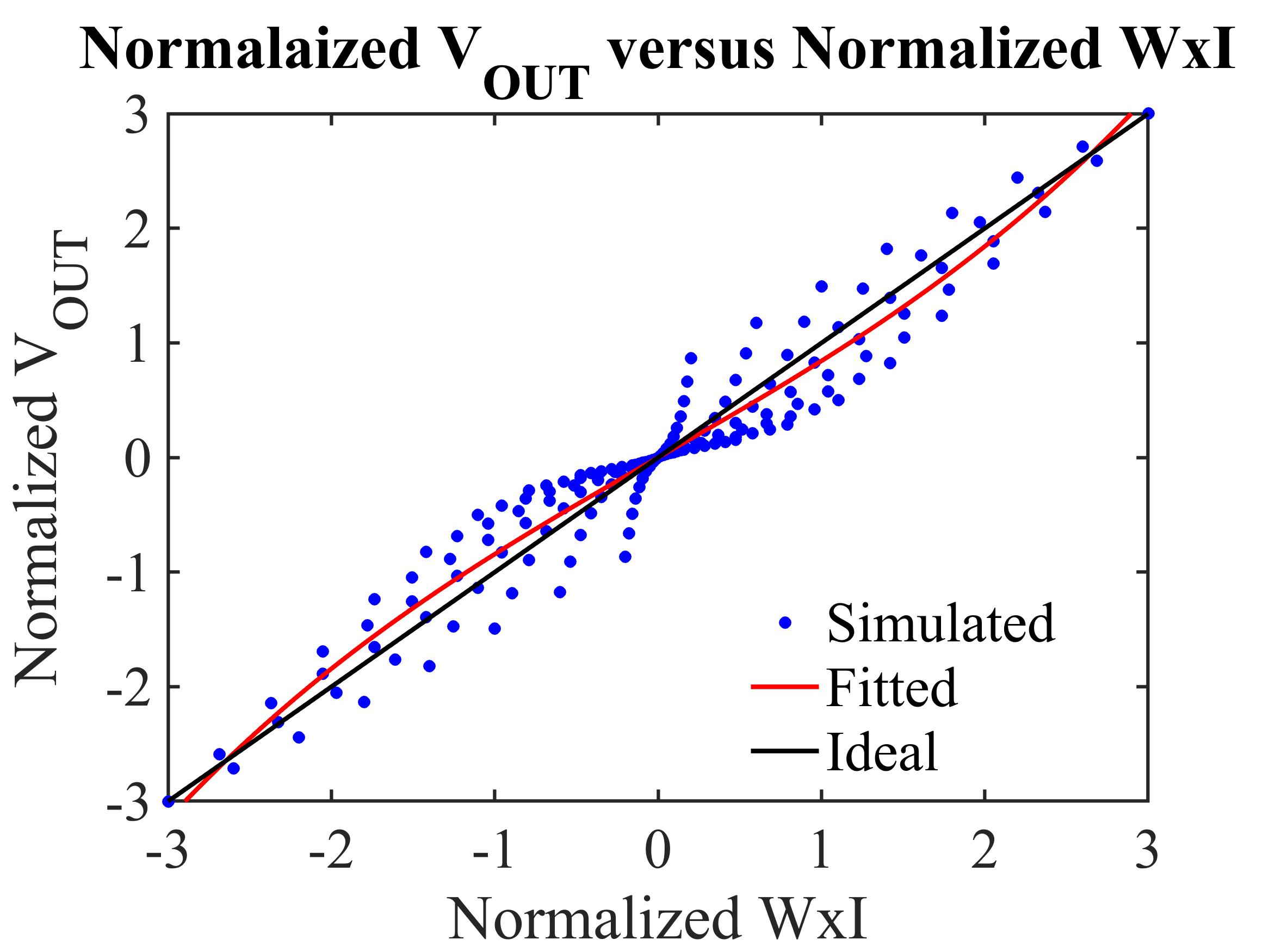}}
\caption{RRAM-based circuit techniques and simulated output for P$^2$M-enabled CIS. (a) Weight embedded pixel circuit, and (b) a scatter plot comparing the simulated convolutional results (normalized \si{V_{OUT}}) with ideal convolutional results (normalized weight\si{\times}input, W\si{\times}I) using GF 22nm FD-SOI process node for a kernel size of 3$\times$3$\times$3.}
\label{nvm_ckt_tf}
\end{figure}

\begin{figure*}[!b]
\centering
 \subfloat[]{\includegraphics[width=0.5\textwidth]{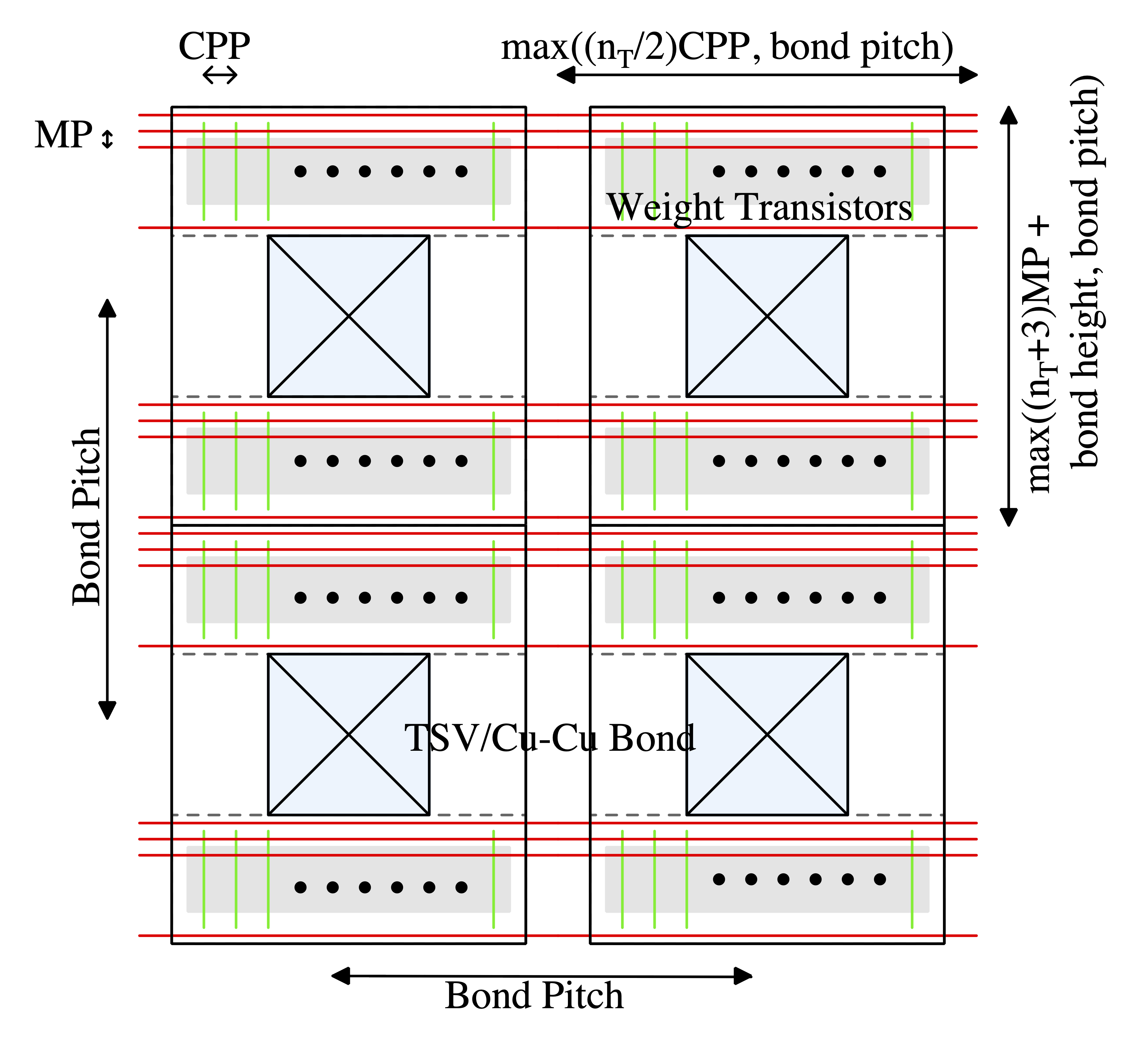}} \\
\subfloat[]{\includegraphics[width=0.5\textwidth]{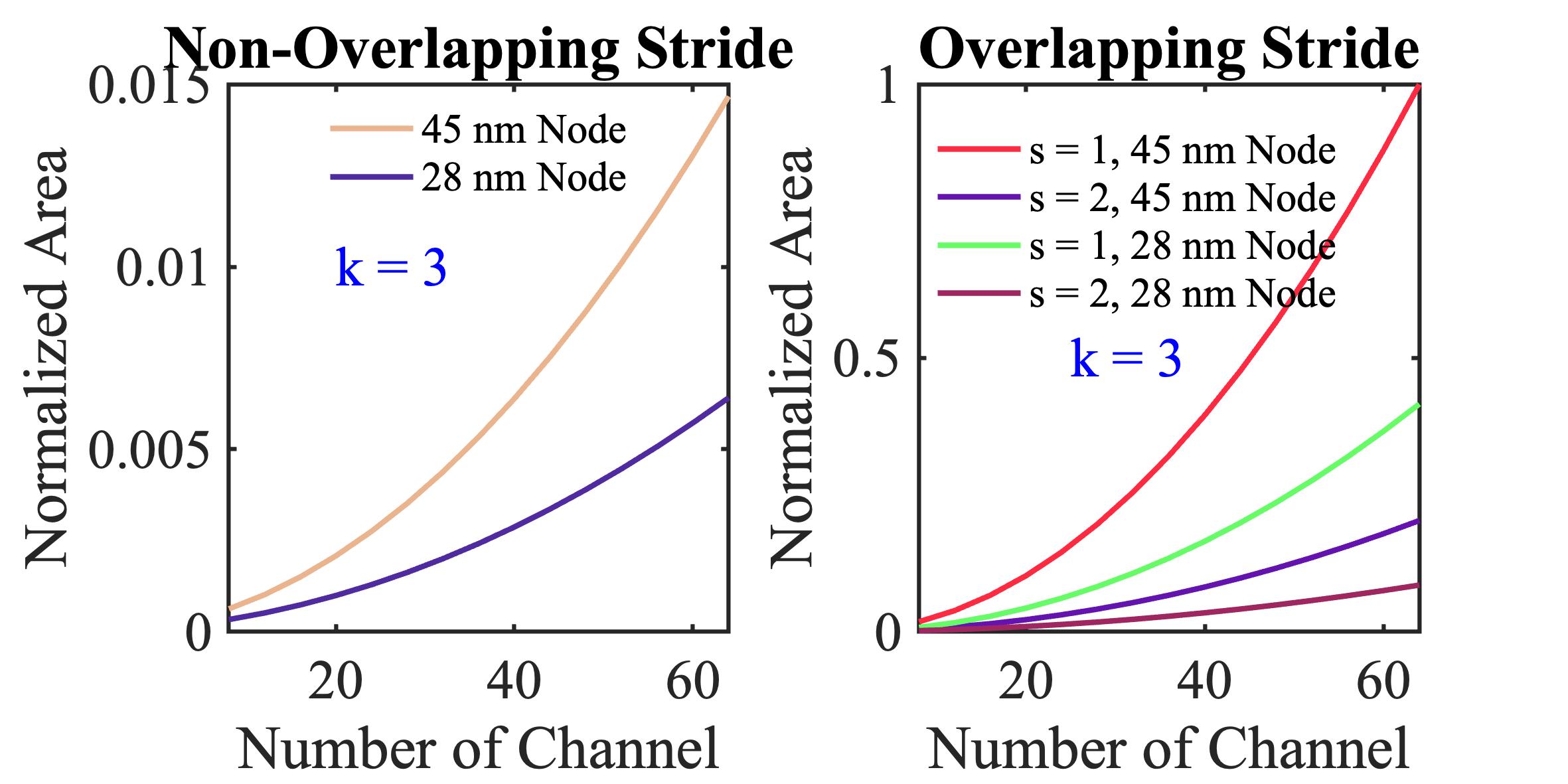}} 
\subfloat[]{\includegraphics[width=0.5\textwidth]{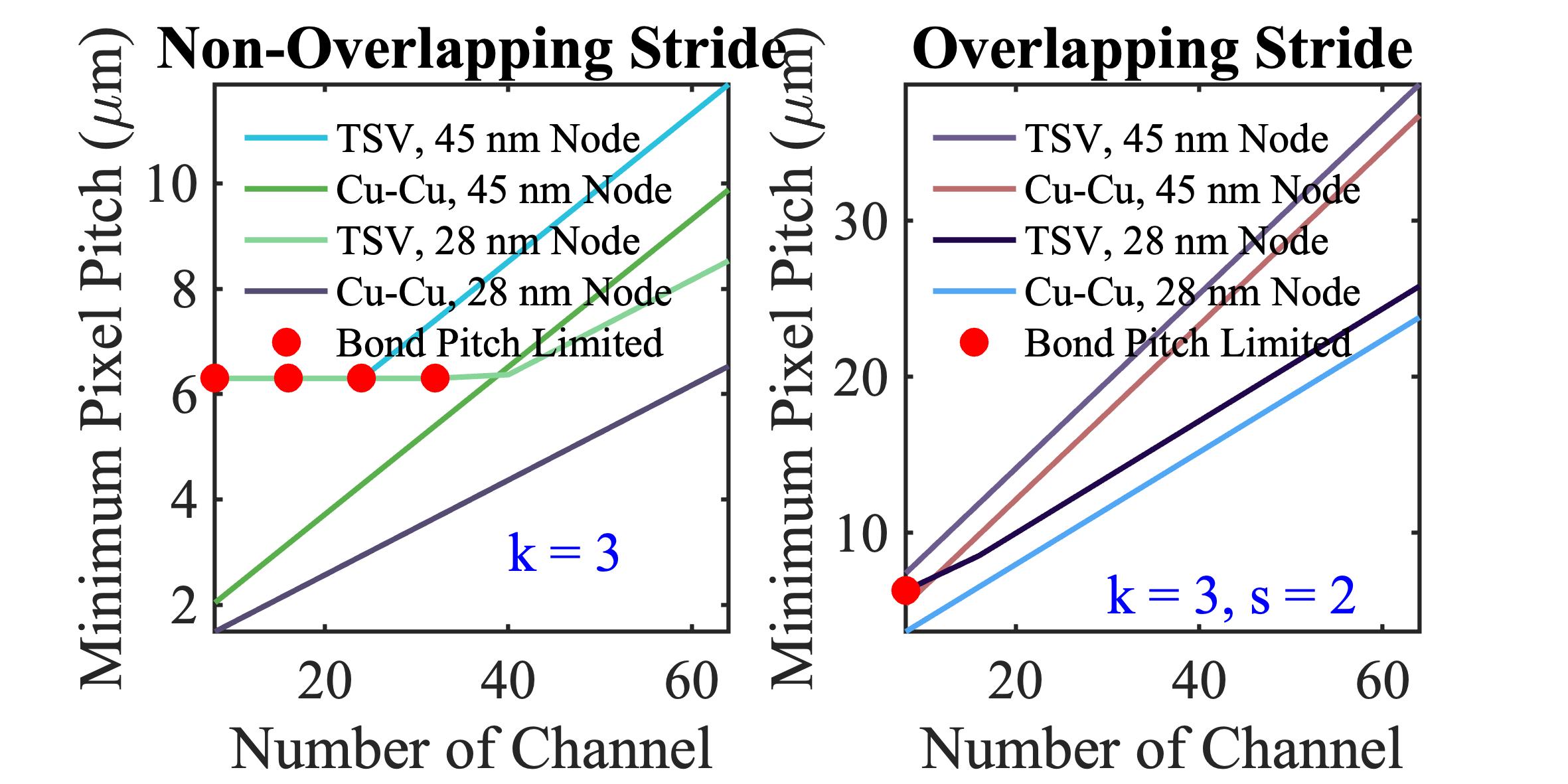}}
\caption{Area trade-off analysis of P$^2$M-enabled CIS. (a) An example layout floor-plan for the weight transistors and 3D integrated bonds, (b) normalized area versus output channels for different process nodes and stride numbers considering Cu-Cu hybrid bonding interface, and (c) minimum pixel pitch versus output channels for different nodes and integration technologies.}
\label{area_figures}
\end{figure*}

\section{Processing-in-Pixel Pre-requisites}

\subsection{Prior Work}

In the first layer of any convolutional neural network (CNN), pixel outputs from the camera are multiplied by multi-bit weight values \cite{algo_channel}. Hence,  the P$^2$M paradigm requires embedding the weights inside the pixel array to enable this computation without sacrificing the CIS resolution. This can be achieved by stacking the weights vertically utilizing different 3D integration technologies \cite{samsung_3D, sony_3D}. Weights can be mapped into the CMOS transistor's geometry or the resistance state of non-volatile memory (NVM) devices, including Resistive Random Access Memory (RRAM), Phase Change Memory (PCM), and Magnetic Random Access Memory (MRAM) \cite{pip_mram}. Moreover, multiple weights must be connected per pixel to support multi-channel convolution operation. Furthermore, weight values can be positive and negative to ensure good test accuracy; hence, circuit techniques are required to differentiate and evaluate positive and negative weights. In \cite{aps_p2m}, the authors leverage the peripheral Single-Slope (SS) ADC to add the MAC results of the positive and negative weights by up-counting and down-counting the counter, respectively, to estimate the final convolution output. These authors re-purpose the on-chip correlated double sampling circuit (CDS) present in CIS and the SS ADC to implement the ReLU operation that introduces non-linear activation required in CNNs by ensuring that the final count value latched from the ADC (after the CDS operation consisting of ‘up’ counting and then ‘down’ counting’) is either positive or zero. These authors also fuse the batch normalization (BN) layer, which is often required for training convergence, partly with the convolutional and partly with the ReLU layer; the BN offset term is implemented by setting an initial value on the counter, and the scale term is multiplied with the weights. Finally, the activations can be transmitted off-chip utilizing different IO (e.g., low voltage differential signaling (LVDS), Interposer (2.5D), through-silicon via (TSV), Wireless, etc.). Fig. \ref{overall_system} represents a generic P$^2$M enabled-CIS system.

\subsection{Proposed RRAM-based in-Pixel Processing}

In this article, we present a CMOS and RRAM-based hybrid implementation of P$^2$M hardware, where we encode the weight values as the resistance states, as illustrated in Fig. \ref{nvm_ckt_tf}(a). The pixel output voltage can be modulated by the gate-source voltage of the source-follower transistor (\si{M_{SF}}) and source-degeneration due to the voltage drop across the memristor (mapped as the weight of a kernel). Multiple RRAMs (\si{M_{Wi}}, where, i = 1 to \si{N_C}) connected at the node "S" of each pixel circuit represent the weights of the different kernels in the output feature map of the first convolutional layer. Each RRAM can be activated individually by the series-connected transistor (\si{M_{EN}}). We performed the convolution operations by activating multiple rows (depending on the layer configurations) and connecting the associated bitlines inside the pixel array for each channel sequentially. The read operation details, BN, and ReLU implementations using Single Slope (SS) ADC can be found in \cite{aps_p2m}. Fig. \ref{nvm_ckt_tf}(b) illustrates the normalized simulated convolutional outputs versus ideal convolutional results using GF 22nm FD-SOI node. A wide range of weight (resistance state) and input light intensity (photodiode current) values for a kernel size of 3\si{\times}3\si{\times}3 have been considered to generate the scatter plot shown in Fig. \ref{nvm_ckt_tf}(b). This simulation uses the resistances of 8 M\si{\Omega} to 23.4 M\si{\Omega}, which may also require device optimization. The simulated results (solid red line, fitted) deviate from the ideal convolutional results (solid black line) due to the circuit's non-linearity, which must be included in the training algorithm by replacing the ideal convolution function with a non-linear custom convolution function.

\section{Technology-Circuit-Algorithm Trade-off Analysis}

\subsection{Area Trade-off Analysis}

The weights are implemented in a separate die heterogeneously integrated with the back-side-illuminated (BI) CMOS Image Sensor (CIS) die. The weights of each pixel need to be stacked vertically and aligned with the pixel pitch to achieve no area overhead. The maximum number of required weight transistors per pixel (\si{n_T}) for strided convolution can be calculated using Eq. \ref{eq_nT} where $k$, $s$, and $c_o$ denote the kernel size, stride, and the number of output channels of the in-pixel convolution layer, respectively \cite{aps_p2m_detrack}. Hence, kernel size, the number of strides, and output channels pose constraints on the minimum pixel pitch. Moreover, the 3D integration technology (e.g., TSV, Cu-to-Cu hybrid bonding, etc.) has a minimum bond pitch requirement that also plays a major role in high-resolution P$^2$M-enabled cameras. An example layout floor-plan of the weight transistors and 3D integrated bonds is shown in Fig. \ref{area_figures}(a). 
The width (\si{w_{PX}}) and height (\si{h_{PX}}) of weight transistors and 3D integrated bonds per pixel can be calculated using Eq. \ref{eq_wpx} and Eq. \ref{eq_hpx}. Here, CPP and MP are process-dependent parameters (45 nm Node: CPP = 190 nm, MP = 140 nm, and 28 nm Node: CPP = 120 nm, MP = 90 nm \cite{wikichip}); on the other hand, $h_{BOND}$ (bond height) and $p_{BOND}$ (bond pitch) are the integration technology dependent parameters. It can be observed from the equation that the minimum pixel pitch of the P$^2$M-enabled CIS can be limited by the weight transistors area (when the number of channels and kernel size is large; however, the stride is small) or the bond pitch, whichever is larger. In summary, design trade-offs exist among CIS pixel pitch, algorithmic parameters ($k$, $s$, and $c_o$), and 3D integration technology. 

\begin{align}
    n_T & = c_o \times \lceil \frac{k}{s} \rceil ^2 \label{eq_nT} \\
    w_{PX} & = max(\frac{n_T}{2} \times CPP, p_{BOND}) \label{eq_wpx} \\
    h_{PX} & = max((n_T + 3) \times MP + h_{BOND}, p_{BOND}) \label{eq_hpx} 
\end{align}

Fig. \ref{area_figures}(b) illustrates the normalized area versus the number of channels for different process nodes considering non-overlapping (left) and overlapping (right) strides for a fixed kernel size of 3, considering Cu-Cu hybrid bondings ($p_{BOND}$ = 1 \si{\mu m}, $h_{BOND}$ = 0.5 \si{\mu m} \cite{cu2cu_pitch}) between the BI-CIS and weight substrates. Due to the finer pitch of the Cu-Cu hybrid bondings, the weight die area for each pixel is determined by the number of weight transistors. As a result, the area becomes a quadratic function of the number of output channels and strides, which can be clearly observed from Fig. \ref{area_figures}(b). The normalized area for a non-overlapping stride for any process node is smaller than the overlapping stride due to fewer weight transistors per pixel. However, we need a lower stride number for better accuracy in our algorithmic model for complex CV tasks; hence, one can use an advanced process node (e.g., 28 nm) for the weight transistors to support more transistors per pixel to achieve low area overhead. For example, it can be observed from Fig. \ref{area_figures}(b) (right) that the stride of 1 for the 28 nm process node incurs \si{\sim}2.4\si{\times } lower area than the 45 nm process node with the same stride number. Note weight transistors from an advanced node should be voltage compatible (e.g., thick oxide device) with the CIS node to achieve sufficient dynamic range.

Fig. \ref{area_figures}(c) represents the minimum pixel pitch versus the number of output channels for different process nodes and integration technologies (TSV: $p_{BOND}$ = 6.3 \si{\mu m}, $h_{BOND}$ = 2.5 \si{\mu m} (diameter) \cite{tsv_pitch}, and Cu-Cu). The bond pitch for TSV is larger than the Cu-Cu pitch; hence, for the advanced process nodes, the minimum pixel pitch can be limited by the bond pitch instead of the number of weight transistors. 
From Fig. \ref{area_figures}(c), it can be observed that, for TSV with a small number of output channels, the minimum pixel pitch is limited by the bond pitch rather than the weight transistors (shown using solid red circles). For instance, the minimum pixel pitch is limited by the bond pitch (= 6.3 \si{\mu}m) for the 28 nm process node up to 32 channels when TSV is used as the bonding interface considering non-overlapping stride (Fig. \ref{area_figures}(c) left). Hence, choosing an advanced process node to support more weight transistors (a large number of channels and a small stride number in favor of the algorithmic accuracy) may also pose minimum pixel pitch constraints unless integration technology with finer bond pitch is also utilized. 

\begin{figure}[!b]
\centering
\includegraphics[width=0.8\linewidth]{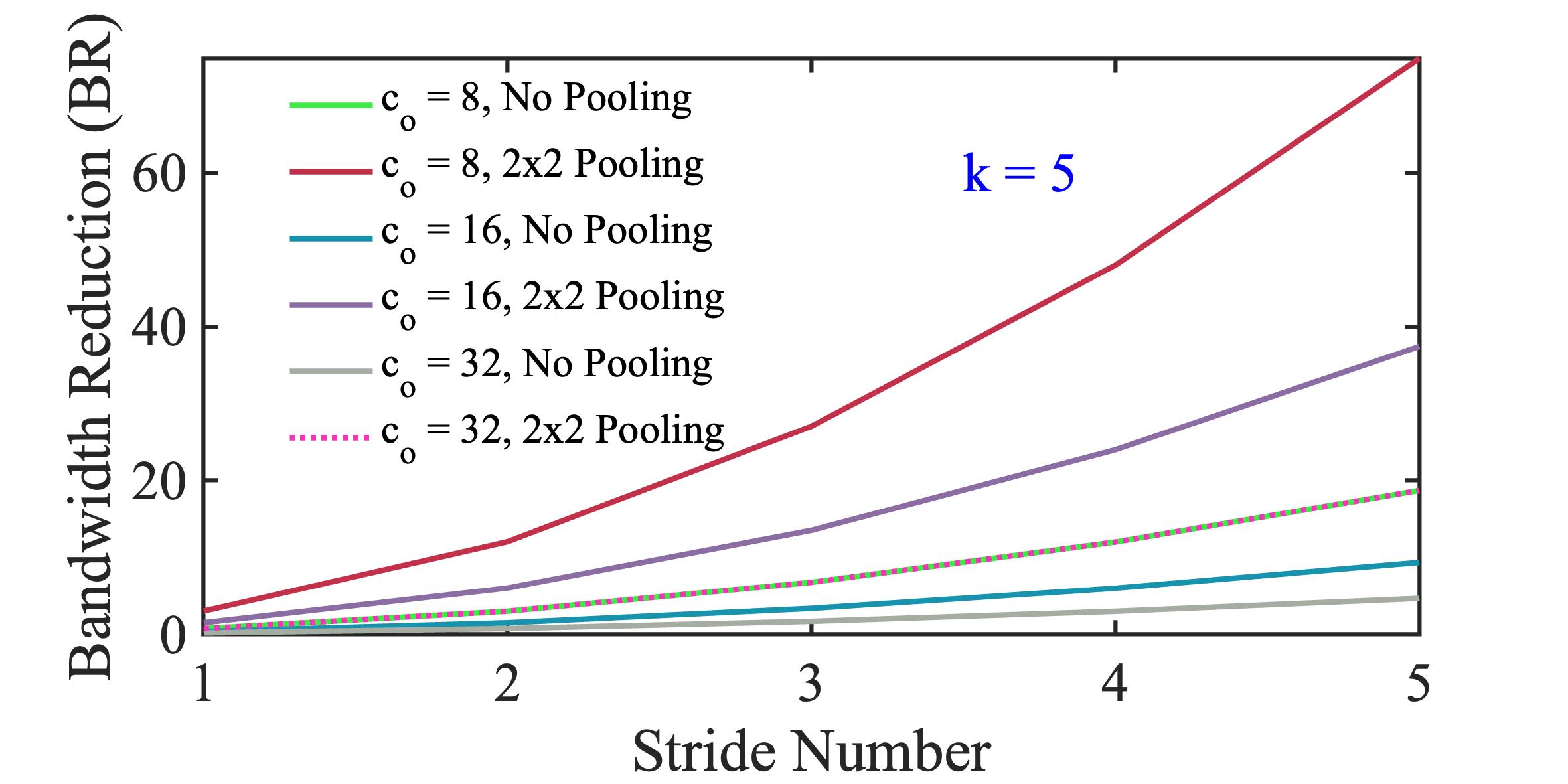}
\caption{Bandwidth reduction (BR) trade-off analysis of P$^2$M-enabled CIS. BR versus stride number for different numbers of output channels and pooling stride.}
\label{br_figures}
\end{figure}

\subsection{Bandwidth Trade-off Analysis}

The data bandwidth reduction (BR) for P$^2$M-enabled CIS can be estimated using Eq. \ref{BR} \cite{aps_p2m}. Here, $I$ and $O$ denote the number of elements of the input RGB image and output activation map of the first convolutional layer in the P$^2$M-enabled solution. $I$ can be calculated as $h_i \times w_i \times 3$, where $h_i$ and $w_i$ denote the height and width dimension of the input image, and factor 3 comes from the RGB channel. $O$ can be calculated using Eq. \ref{O} and Eq. \ref{ho_wo}, where $h_o$, $w_o$, and $p$ represent the height and width of the output activation map and padding, respectively. The $\frac{4}{3}$ factor represents the compression from Bayer's RGGB pattern to RGB pixels, and $b_{ADC}$ denotes the ADC bit precision ($b_{ADC}$ = 8 is used here) in our P$^2$M approach.

\begin{align}
    BR & = \Big(\frac{I}{O}\Big) \Big(\frac{4}{3}\Big) \Big(\frac{12}{b_{ADC}}\Big) \label{BR} \\
    O & =  h_o \times w_o \times c_o \label{O} \\
    h_o (w_o) & = \frac{h_i(w_i) - k + 2\times p}{s} + 1 \label{ho_wo} 
\end{align}

\begin{figure}[!b]
\centering
\includegraphics[width=0.8\linewidth]{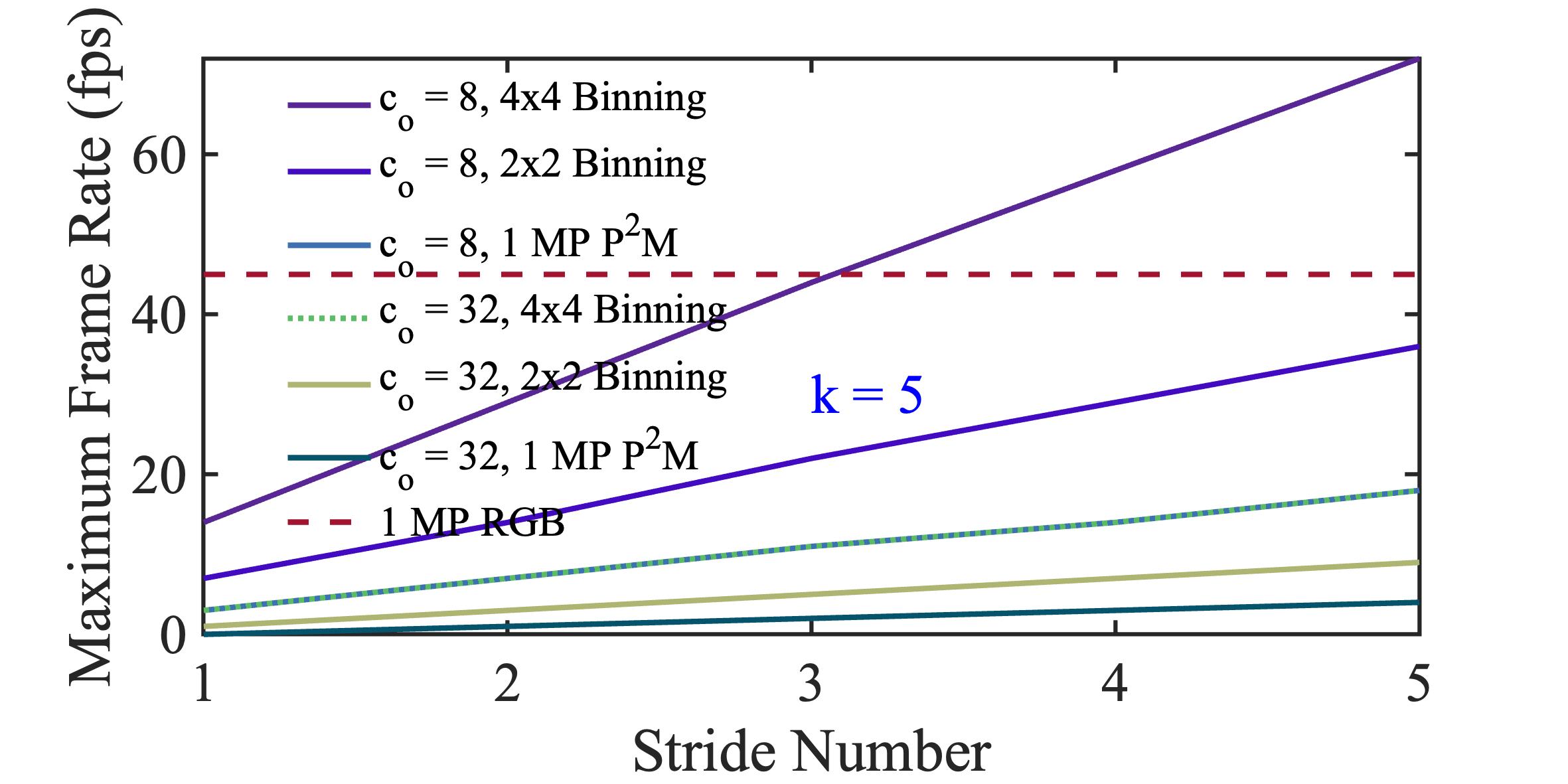}
\caption{Latency trade-off analysis of P$^2$M-enabled CIS. Maximum frame rate versus stride number for different numbers of output channels and pixel binning.}
\label{latency_figures}
\end{figure}

Fig. \ref{br_figures} illustrates the estimated BR versus stride number for a kernel size of 5, considering  a different number of output channels and pooling. P$^2$M approach can provide a significant amount of data bandwidth reduction compared to conventional CIS when the stride is large, and the number of output channels is small. Communication bandwidth becomes more important for a system when transmission cost dominates (e.g., wireless short/long range); hence, algorithm optimization (e.g., choosing a non-overlapping stride with few output channels while maintaining accuracy requirement) is necessary. 
Moreover, a better BR number can be achieved by utilizing pooling (max/average) in the periphery after the output of the first convolution layer of the P$^2$M approach.

\subsection{Latency Trade-off Analysis}

The frontend latency depends on the number of read cycles ($n_C$) calculated using Eq. \ref{n_read} for our P$^2$M approach. The total frontend latency ($T_{FRONTEND}$) can be estimated using Eq. \ref{Td}, where $T_{EXP}$, $T_{ADC}$, and $T_{IO}$ represent the exposure time, ADC read time and communication delay associated with IO pins. The IO delay ($T_{IO}$) depends on ADC bit precision ($b_{ADC}$), IO bandwidth ($BW_{IO}$) (e.g., LVDS: 1 Gbps \cite{lvds_io}), and the total number of IO pads ($n_{IO(PAD)}$) used on the chip. Fig. \ref{latency_figures} illustrates the maximum frame rate ($\frac{1}{T_{FRONTEND}}$) for our P$^2$M-enabled CIS versus stride number for different output channels and pixel binning, considering a kernel size of 5. It can be observed from the figure that the frontend maximum frame rate (FR) of the P$^2$M approach is smaller than conventional RGB CIS for most cases. The total number of read cycles ($n_C$) increases with the number of output channels. Moreover, the exposure and the ADC read latencies are twice in our P$^2$M solution compared to conventional CIS due to considering positive and negative weights inside the first convolutional layer. However, by utilizing a large non-overlapping stride, a small number of output channels, and pixel-level binning (by reducing the $n_C$), a high frame rate can be achieved ($c_o = 8$, $4 \times 4$ binning for stride number = 5 shown in the figure); nevertheless, it may affect the accuracy. 

\begin{align}
    n_{C} & = h_o \times c_o \label{n_read} \\
    T_{FRONTEND} & = n_{C} \times (T_{EXP} + T_{ADC} + T_{IO}) \label{Td} \\
    T_{IO} & = \frac{w_o \times b_{ADC}}{BW_{IO} \times n_{IO(PAD)}} \label{Tio} 
\end{align}

Note for in-pixel computing; we define the frontend latency as the total time to compute all activations of the output feature map  ($h_o \times w_o \times c_o$) for the first convolution layer. As a result, the latency values to compute that many elements are typically higher than conventional CIS latency unless there is a large suppression in the output feature map (e.g., few channels, large stride, binning). However, for non-overlapping strides, the frame rate can be boosted by utilizing the idle peripheral ADC with a cost of multiple bitlines. For instance, only one ADC is required to compute the convolution output of a kernel size of $k \times k$, while ($k-1$) ADC remains idle. Hence, adding extra $k$ bitlines for non-overlapping stride can improve the latency by $k$ times utilizing $k$ ADCs in parallel \cite{aps_p2m_detrack}. This parallelism can be easily implemented for non-overlapping strides; however, overlapping strides require complex bitline structure and control circuitry. Note the absolute value of the FR can also be improved by reducing exposure time, independent of underlying P$^2$M schemes.    

\begin{figure}[!b]
\centering
\includegraphics[width=0.8\linewidth]{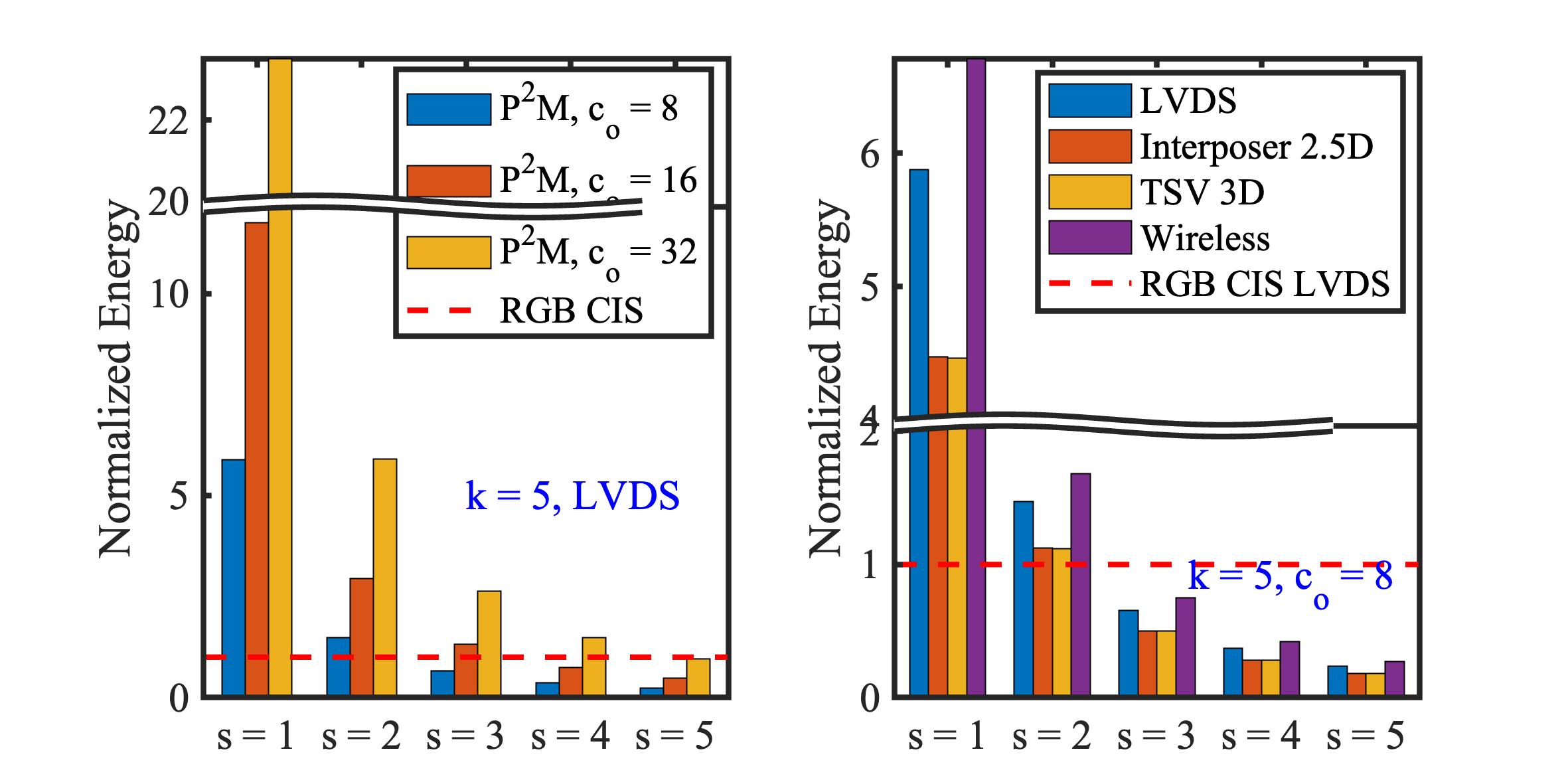}
\caption{Frontend energy trade-off analysis. (Left) Normalized energy versus stride for different numbers of output channels considering LVDS IO, and (right) normalized energy versus stride for different IO configurations.}
\label{energy_figures}
\end{figure}

\subsection{Energy Trade-off Analysis}

The total energy for P$^2$M-enabled CIS depends on the number of total convolution operations ($n_{READ}$) that can be calculated using Eq. \ref{n_conv}. 
It can be deduced from the equation that a large stride and few channels can help to achieve lower energy. Moreover, we can use lower ADC bit precision (e.g., 8 bits) than typical 12-bit precision for conventional pixel readout. The total frontend energy can be calculated using Eq. \ref{energy}, where $e_{PX}$, $e_{ADC}$, and  $E_{IO}$ denote the convolutional energy and ADC read energy per operation, and total communication (IO) energy, respectively. The IO energy can be estimated using Eq. \ref{io_energy}, where $e_{IO}$ denotes the energy per bit for different IO technologies (LVDS: 12.34 pJ/bit \cite{lvds_io}, Interposer (2.5D): 259.9 fJ/bit \cite{io_energy}, TSV (3D): 176.2 fJ/bit \cite{io_energy}, Wireless (WiFi): 19.5 pJ/bit \cite{wifi_energy}). 

\begin{align}
    n_{READ} & = h_o \times w_o \times c_o \label{n_conv} \\
    E_{FRONTEND} & = n_{READ} \times (e_{PX} + e_{ADC}) + E_{IO}  \label{energy} \\
    E_{IO} & = h_o \times w_o \times c_o \times b_{ADC} \times e_{IO} \label{io_energy}
\end{align}

Fig. \ref{energy_figures} illustrates the normalized energy versus different stride numbers considering different output channels (left) and different IO technologies (right), considering a kernel size of 5. It can be observed from the left subplot of the figure that non-overlapping stride (s = 5) provides energy benefit for $c_o = 8, 16,$ and $32$.  However, more channels (e.g., 64) in the first convolutional layer of P$^2$M may not provide any benefit due to the increasing number of total convolution operations ($n_{READ}$). Moreover, for overlapping strides (e.g., s = 2), the P$^2$M may consume more energy than the conventional RGB CIS system due to the increasing size of the output feature map ($h_o$ and $w_o$). Note this highlights the fact that P$^2$M is inherently a hardware-algorithm co-design problem, and merely implementing computing inside the pixel array does not guarantee energy benefit without algorithm optimization. 
In addition, the normalized energy depends on different IO technologies (e.g., LVDS, Interposer, TSV, WiFi), which can be observed from the figure (right). TSV (3D) and Interposer (2.5D) consume lower energy from the LVDS and Wireless (WiFi) transmission due to lower link parasitics. 

\subsection{Accuracy Trade-off Analysis}

\begin{figure}[!t]
\centering
\includegraphics[width=0.8\linewidth]{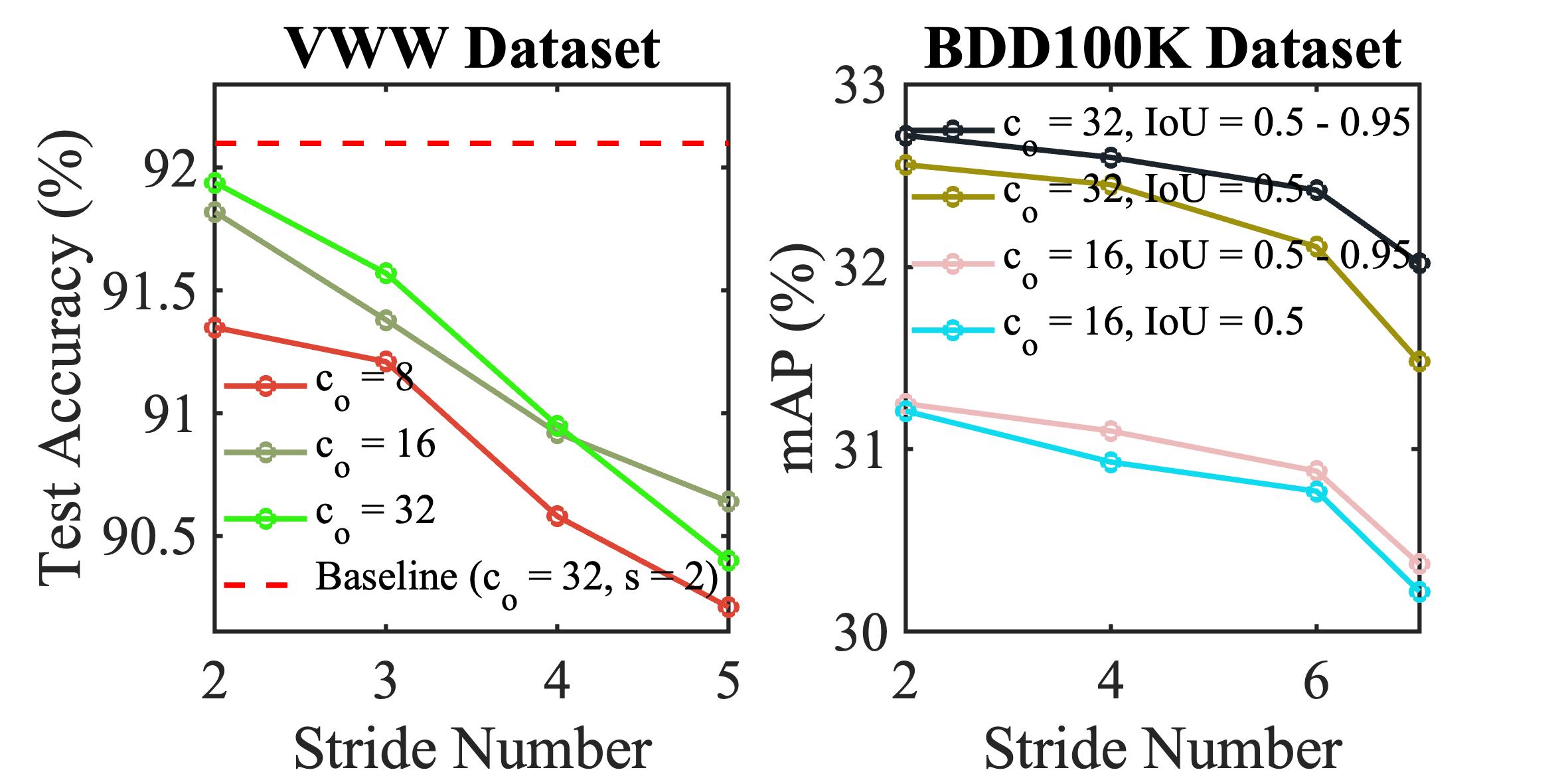}
\caption{Accuracy trade-off analysis of P$^2$M-enabled CIS. (a) visual wake word (VVW) dataset, and (b) BDD100K dataset.}
\label{accuracy_figures}
\end{figure}

The accuracy with P$^2$M-implemented systems can be largely maintained with noise-aware training where the non-linearity is extracted as a curve-fitted function that replaces the traditional convolution function in the ML framework. However, the accuracy is impacted by the hardware constraints that motivate low bit-width, large strides, and few filters, i.e., the layer configuration that leads to large dimensionality reduction. The impact is larger with complex applications (e.g. multi-object tracking from the BDD100K dataset \cite{bdd100k}) that demand more fine-grained feature extraction and require small dimensionality reduction. As shown in Fig. \ref{accuracy_figures}, increasing the stride from 2 to 5 with 16 channels decreases the mean average precision by $2.8\%$ on BDD100K. On the other hand, a higher stride increase from 2 to 6 with the same 16 channels degrades the test accuracy of the visual wake words dataset only by $1.28\%$. A similar trend is also observed with the number of filters. Hence, our P$^2$M paradigm is more attractive to on-device edge workloads with tight compute and memory budgets that also demand large BR.  

\section{Future Directions}

\subsection{Improved Non-Linearity Modeling for P$^2$M} 

Although P$^2$M has shown sufficient promise to accelerate deep learning (DL), significant challenges exist, including non-deterministic noise and non-linearity, that lead to significant accuracy drops for complex CV tasks \cite{aps_p2m_detrack}. In particular, the non-linearity-induced accuracy drop is due to the mismatch between the accurate SPICE simulations that capture the interactions of non-linear devices and their simplified functional representations that can be readily incorporated in ML frameworks (e.g. PyTorch). To yield SOTA accuracies with analog P$^2$M circuits, we have to 1) identify classes of non-linear functions that are general enough to capture analog behaviors but simple enough to train complex DL models and 2) train DL models with these functions and the non-deterministic noise.

\subsection{Frame Skipping}

Skipping frames based on their relative importance to the accuracy of the downstream task can reduce the sensor and the overall system energy consumption by (approximately) the rate of frame skipping, assuming the associated overhead is manageable \cite{Ghodrati2021FrameExitCE}. For complex CV tasks such as multi-object tracking from BDD100K, our initial results indicate that alternate frames can be skipped naively without any impact on the final accuracy. This implies there are significant redundancies in the scenes captured by the sensors. This motivates the development of training frameworks that learn to skip frames more aggressively, processing fewer frames for simpler scenes and more for more complex scenarios so there is no performance degradation. In addition, skipping regions of frames that contain no significant change from the previous frame will yield further energy benefits.

\subsection{Distributed Computing and Sensor Fusion}

Embedding only a few layers in the sensor chip can result in only marginal energy and cost savings (from a manufacturing point of view), especially for a deep network (e.g. ResNet152). This can be largely mitigated with a distributed computing paradigm where the DL computation is split between the pixel chip that processes the first few layers as illustrated in P$^2$M, a logic chip heterogeneously (e.g. monolithic 3D or $\mu$TSV that consumes significantly lower energy compared to standard camera interfaces) 
integrated with the sensor that processes some intermediate layers (number of layers subject to peak memory and compute constraints), and off-chip hardware, such as FPGA, that processes the remaining layers \cite{sony_3D}. We can process many more layers in the logic chip than the pixel chip, which can significantly reduce the spatial dimensions of the network, thereby reducing the data bandwidth and alleviating the compute burden of the off-chip hardware. 
The optimal network configuration and the splitting points, subject to each hardware's compute and memory constraints, can be obtained from a SuperNet-based neural architecture search (NAS) framework. 
Lastly, P$^2$M is an attractive hardware platform for sensor-fusion applications (e.g. autonomous driving and satellite imaging), where the first few/several layers of each feature extraction network can be embedded inside each heterogeneously integrated sensor chip. This will broaden the energy efficiency promise of in-sensor computing.

\section{Discussions and Conclusions}

In this article, we analyze and present solutions to mitigate the challenges in existing P$^2$M paradigms that aim to embed CNN computations inside the pixel array of traditional CIS. Such challenges primarily stem from image sensors being optimized for reading out pixels and not processing computational requirements of modern CNN layers, including multi-bit, multi-channel strided convolution with both positive and negative weights, BN, ReLU/LIF, and pooling. We propose a technology-circuit-algorithm tri-design solution that resolves the conflicting requirements emerging from the technology and algorithmic layers; technology constraints require large strides and a reduced number of channels for fewer weight transistors to meet the pixel density, while the algorithmic constraints require fewer strides and increased number of channels to meet accuracy requirements, particularly for complex applications. Unlike existing P$^2$M works that leverage the driving strength of CMOS transistors to implement weights, we also present NVM-based P$^2$M solutions where weights can be reconfigurable, yielding real-time adaptation capabilities. This study can be extended to \textit{in-sensor} and \textit{near-sensor} processing paradigms with similar conflicting requirements.
\textit{A critical takeaway from this paper is the fact that in-pixel processing is not merely about adding relevant machine learning computations inside the pixel array; it is inherently a co-design problem where the solution spaces have to be optimized both on the circuit as well as on the algorithm side simultaneously to achieve significant benefit in power, performance, area, bandwidth reduction, and accuracy}.

\bibliographystyle{unsrtnat}
\bibliography{references}

\end{document}